\documentclass[aps,amsmath,amssymb,twocolumn]{revtex4-1}

\usepackage{epsfig}
\usepackage{graphicx}
\usepackage{graphicx}
\usepackage{dcolumn}
\usepackage{bm}
\usepackage{setspace}
\usepackage{amsmath}
\usepackage{array}
\usepackage{natbib}
\usepackage{color}

\begin{document}

\title{Capillary fracture of soft gels}
\author{Joshua B. Bostwick}
\affiliation{Department of Mathematics, North Carolina State University, Raleigh, NC 27695, USA}
\author{Karen E. Daniels}
\affiliation{Department of Physics, North Carolina State University, Raleigh, NC 27695, USA}


\begin{abstract}
A liquid droplet resting on a soft gel substrate can deform that substrate to the point of material failure, whereby fractures develop on the gel surface that propagate outwards from the contact-line in a starburst pattern.
In this paper, we characterize i) the initiation process in which the number of arms in the starburst is
controlled by the ratio of surface tension contrast to the gel's elastic
modulus and ii) the propagation dynamics showing that once fractures are initiated they propagate with a universal power  law $L\propto t^{3/4}$. We develop a model for crack initiation by treating the gel as a linear elastic solid and computing the deformations within
the substrate from the liquid/solid wetting forces.
The elastic solution shows that both the location and magnitude of the wetting forces are critical in providing a quantitative prediction for the number of
fractures and, hence, an interpretation of the initiation of {\itshape capillary
fractures}. This solution also reveals that the depth of the gel is an important factor in
the fracture process, as it can help mitigate large surface tractions; this
finding is confirmed with experiments. We then develop a model for crack propagation by considering the transport of an
inviscid fluid into the fracture tip of an incompressible material, and find
that a simple energy-conservation
argument can explain the observed material-independent power law. We compare predictions for both linear elastic and
neo-Hookean solids finding that the latter better explains the observed exponent.
\end{abstract}

\maketitle


\section{Introduction}

The interaction of soft substrates with fluid interfaces is common to many biological, medical, and industrial processes. For this reason, the field of elastocapillarity \cite{Roman2010}, in which one studies how surface tension forces couple to the deformations of elastic substrates, has been the subject of much recent attention. Even the most basic characterization of the wetting forces at the three-phase contact-line remains in dispute, especially for soft viscoelastic materials \cite{kajiya13} that have both liquid and solid properties. For such materials, the Young-Dupr\'{e} law for a liquid wetting a hard solid, Neumann's law for a liquid wetting another liquid, or some hybrid thereof may apply \cite{Snoeijer,Weijs2013}.
Theoretical efforts have been made to bridge the gap between the hard solid and liquid regime, such as introducing the concept of solid surface tension \cite{jerison11,style12}. Alternative methods employ computational approaches such as density functional theory (DFT) \cite{Das2011} and molecular dynamic (MD) simulations \cite{Weijs2013} to gain a more thorough understanding of the wetting forces acting at the contact-line.

In general, the elastic resistance must be comparable in magnitude to the surface tension forces applied to the elastic medium for many elastocapillary phenomenon. The elastocapillary number $\sigma/(EL)$ is a typical measure of the relative importance of capillarity to elasticity. Here $\sigma$ is the surface tension, $E$ the elastic modulus and $L$ a characteristic length scale.
In experiments, it is typically easiest to adjust $L$, as seen in the wrinkling of elastic sheets \cite{huang07,vella10,Davidovitch2011}, capillary origami \cite{jung09} and buckling of elastic fibers \cite{evans13}.
However, it is also possible to use a gel as the solid phase \cite{Mora2006,daniels07,jerison11,kajiya13}, which permits $E$ to be tuned over several orders of magnitude.

\begin{figure}
\begin{center}
\begin{tabular}{cc}
($a$) & ($b$)  \\
\includegraphics[width=0.25\textwidth]{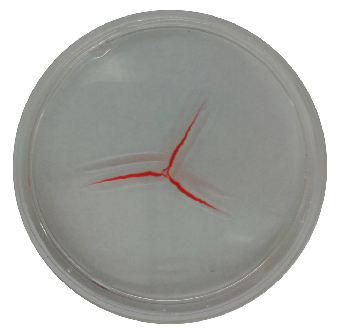} &
\includegraphics[width=0.24\textwidth]{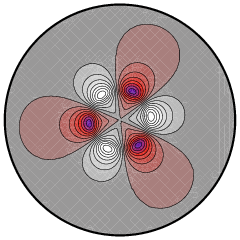}
\end{tabular}
\end{center}
\caption{\label{fig:intro} (Color online) ($a$) A starburst fracture, formed by placing a droplet of liquid at the center of a Petri dish containing agarose gel. ($b$) Computed displacement field, with $n=3$ arms, induced by the wetting forces from the droplet. }
\end{figure}

We focus our attention on recent experiments in which surfactant-laden fluid droplets are applied to soft gel substrates \citep{daniels07,spandagos12a,spandagos12b}, as shown in Figure~\ref{fig:intro}. In these experiments, a partially-wetting liquid spreading on an agarose or gelatin substrate can produce \emph{capillary fractures} that originate at the contact-line and propagate outwards in a starburst pattern. It is convenient to divide the fracture process into three phases: i) initiation, ii) nucleation and iii) propagation.
The initiation phase sets a critical wavenumber through the elastic deformation field that controls the number of arms within the starburst fracture.
In previous experiments, the number of arms has been shown to be controlled by the ratio of surface tension contrast to the gel's elastic modulus \citep{daniels07}.
The nucleation process in soft materials such as agarose gel is not deterministic, rather thermal fluctuations begin the fracture dynamics after some finite time delay \citep{bonn,Wang2012a}. Once the cracks are nucleated, they fill with fluid from the droplet and grow with a universal power-law $L\propto t^{3/4}$  which does not scale with any material parameters\citep{daniels07}. Intriguingly, even the application of a super-spreading surfactant (Silwet 77) does not increase the exponent \citep{spandagos12a,spandagos12b}. In this paper, we analyze the initiation and propagation processes separately. 

\begin{figure}
\begin{center}
\includegraphics[width=0.3\textwidth]{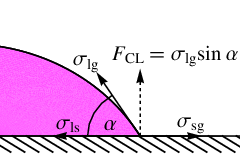}
\end{center}
\caption{\label{fig:YD} (Color online) The Young-Dupr\'{e} equation (\ref{YD}) schematically as a horizontal force balance, as well as the unbalanced (vertical) contact-line force $F_{CL}$.}
\end{figure}
To study problems coupling elasticity to capillarity, it is helpful to put the typical length scales involved with such deformations into perspective. For a liquid on a hard substrate, the wetting properties are defined by the Young-Dupr\'{e} equation \cite{young,dupre},
\begin{equation}
\sigma_{sg}-\sigma_{ls}= \sigma_{lg} \cos \alpha, \label{YD} \end{equation}
which relates the liquid/gas $\sigma_{lg}$, liquid/solid $\sigma_{ls}$ and solid/gas $\sigma_{sg}$ surface tensions to the static contact-angle $\alpha$. Figure~\ref{fig:YD} illustrates the interpretation of the Young-Dupr\'{e} relationship as a horizontal force balance. Note that this formulation also leads to an \emph{imbalance} of vertical forces with magnitude $F_{CL}=\sigma_{lg} \sin\alpha$. For a soft substrate with elastic modulus $E$, this force gives rise to deformations of size $\ell \sim \sigma_{lg}/E$, more commonly referred to as the elastocapillary length.
For example, water ($\sigma_{lg}=72$~mN/m) wetting a glass substrate ($E=70$~GPa) yields a characteristic deformation $\ell \sim 10^{-12}$~m, justifying the neglect of the unbalanced force. Recently, Jerison \emph{et al.} have used fluorescence confocal microscopy to show that water droplets interacting with silicone gel ($E=3$~kPa) yield micron-size substrate deformations $\ell \sim 10^{-6}$~m \cite{jerison11,style12}. For the capillary fracture experiments shown in Figure~\ref{fig:intro}, which involve {\itshape ultra-soft} agaraose substrate ($E\sim 10$~Pa), the elastocapillary length corresponds to $\ell \sim 10^{-3}$~m (millimetric) deformations. Given the size of such deformations, it is therefore unsurprising that fractures occur. 

\begin{figure}
\begin{center}
\begin{tabular}{c}
($a$)  \\
\includegraphics[width=0.4\textwidth]{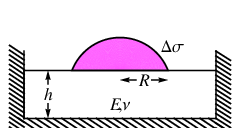} \\ ($b$) \\
\includegraphics[width=0.4\textwidth]{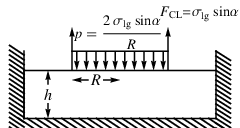}
\end{tabular}
\end{center}
\caption{\label{fig:defsketch} (Color online) Definition sketch: ($a$) schematic of the liquid droplet wetting an elastic substrate and ($b$) the associated wetting forces included in the model. }
\end{figure}

We begin this paper by defining the equations governing the deformation of an elastic substrate loaded by a partially-wetting liquid droplet with a corrugated contact-line in \S\ref{sec:initiation}. We solve the governing equations to show there is a critical disturbance with wavenumber $n_c$ that generates the largest elastic response, as measured by the tangential displacement, which we associate with the number of arms in a starburst fracture. The solution illustrates that the location $R$ of the unbalanced contact-line force is an important parameter in wavenumber selection. This dependence is quantitatively confirmed in experiments where the wetting conditions $\alpha$ are fixed and the volume $V$ is varied, validating our interpretation of the initiation of capillary fractures. In addition, the depth of the gel can play a similar role in the initiation process, which we confirm with experiments. Section \ref{sec:nucleation} briefly discusses the nucleation process. 
In \S\ref{sec:propagation}, we model the crack propagation dynamics by considering the transport of an inviscid fluid into the crack tip of an elastic solid, subject to energy conservation (Bernoulli's law). While it is reasonable to assume that the solid behaves as a linear elastic material for the deformation phase, this assumption is almost certainly violated once a fracture has been nucleated. Therefore, we develop a model for the propagation dynamics which allows for considering either a i) linear elastic or ii) neo-Hookean solid. 
Both elastic models for the solid predict power law growth, but the exponent from the neo-Hookean model matches the observed power law behavior $L(t) \propto t^{3/4}$. The predicted exponent and pre-factor do not depend on any of the material parameters, consistent with experimental observations \cite{daniels07,spandagos12a,spandagos12b}. We conclude with some remarks in \S\ref{sec:discussion} regarding capillary fracture in soft materials.



\section{Crack initiation \label{sec:initiation}}

Consider a liquid droplet, contained at its free surface by the liquid-gas
surface tension $\sigma_{lg}$ and resting on a
linear elastic substrate of thickness $h$, which is characterized by an elastic
modulus $E$ and Poisson ratio $\nu$, as shown in
Figure~\ref{fig:defsketch}($a$).
This partially-wetting liquid interacts with the solid through the
capillary pressure $p=2 \sigma_{lg} \sin \alpha/R$ uniformly distributed over
the liquid/solid contact area and the unbalanced (vertical) contact-line force
$F_{CL}=\sigma_{lg} \sin \alpha$ applied at the contact-line radius $R$
(see Figure~\ref{fig:defsketch} ($b$)). Here we note that an alternative model of wetting which, in addition to the capillary pressure and vertical contact-line force, also includes a horizontal contact-line force directed into the liquid phase has recently been put forth \cite{Snoeijer,Weijs2013}. For the purposes of this paper, however, we restrict ourselves to a model of wetting that includes the capillary pressure and vertical contact-line force. The extent to which the liquid partially wets the solid is controlled by the surface tension contrast $\Delta \sigma =
\sigma_{sg}-\sigma_{ls}$, or equivalently the static contact-angle $\alpha$,
defined by the Young-Dupr\'{e} equation (Equation~\ref{YD}). We characterize
the elastic response in the substrate due to both the capillary pressure and the contact-line force associated with a corrugated contact-line.

\subsection{Field equations}
We begin by introducing the displacement field $\boldsymbol{u}$,
\begin{equation}
\boldsymbol{u}=u_r(r,\theta,z) \boldsymbol{\hat{e}}_r +
u_\theta(r,\theta,z)\boldsymbol{\hat{e}}_{\theta}+u_z(r,\theta,z)
\boldsymbol{\hat{e}}_z ,
\label{dispdef}
\end{equation}
in cylindrical coordinates ($r,\theta,z$), which satisfies the governing
elastostatic Navier equations,
\begin{equation}
\left(1-2\nu\right) \boldsymbol{\nabla}^2 \boldsymbol{u} +
\boldsymbol{\nabla}\left(\boldsymbol{\nabla}\cdot \boldsymbol{u}\right) = 0.
\label{navier}
\end{equation}
The strain field $\boldsymbol{\varepsilon}$ is defined as
\begin{equation}
\boldsymbol{\varepsilon} = \frac{1}{2}
\left(\boldsymbol{\nabla}\boldsymbol{u} +
\left(\boldsymbol{\nabla}\boldsymbol{u}\right)^t\right),
\label{kinematic}
\end{equation}
while the stress field $\sigma_{ij}$ for this linear elastic solid is given by
\begin{equation}
\sigma_{ij} = \frac{E}{1+\nu} \left(\varepsilon_{ij} +
\frac{\nu}{1-2\nu} \,\varepsilon_{kk}\right).
\label{constitutive}
\end{equation}

\subsection{Boundary conditions}
We assume the elastic substrate is pinned to a rigid support at $z=0$ by
enforcing a zero displacement boundary condition there,
\begin{equation}
\boldsymbol{u}(r,\theta,0)=0. \label{dispBC}
\end{equation}
In experiments this support corresponds to the bottom of the dish holding the substrate material. Similarly, we specify the surface tractions on the
free surface $z=h$,
\begin{equation}\begin{split}
&\sigma_{zz}(r,\theta,h) - \Sigma_s
\nabla^2_{\parallel}u_z (r,\theta,h) = F(r,\theta),\; \\
&\sigma_{rz}(r,\theta,h)=\sigma_{\theta z}(r,\theta,h)=0,
\label{forceBC}\end{split}
\end{equation}
where $\nabla^2_{\parallel}$ is the surface Laplacian and $F(r,\theta)$ is
the applied force associated with the liquid/solid (wetting) interactions. Here we introduce the solid surface tension $\Sigma_s$ for rigor as well as to regularize the singularity associated with applying a delta function force to the surface of an elastic medium, as discussed in \cite{style12}.

We now develop a model for the forces $F(r,\theta)$ associated with the
wetting of a
liquid droplet on a soft elastic substrate. First, we note that
the surface of our gel substrate is heterogeneous so that the location of
the contact-line generally takes the form $r=R(\theta)$(in fact, corrugations are typically present even for ideal solid or liquid
substrates).
For a liquid droplet
held by uniform surface tension $\sigma_{lg}$, the wetting forces associated
with this configuration are given by
\begin{equation}
F(r, \theta) = -p \,H\left(R\left(\theta\right)-r\right)+ F_{CL}
\,\delta\left(r-R\left(\theta\right)\right),
\label{forceG}
\end{equation}
where $H$ and $\delta$ are the Heaviside and delta functions, respectively (see
Figure~\ref{fig:defsketch}~($b$)). Here the capillary pressure $p=2 \sigma_{lg} \sin \alpha /R$ is uniformly distributed over the liquid/solid surface area,
whereas the unbalanced contact-line force $F_{CL}=\sigma_{lg} \sin \alpha$ is
applied as a point load at the contact-line. Note the orientation of the applied
forces; the capillary pressure $p$ compresses the substrate, while the
contact-line force $F_{CL}$ tends to pull the substrate upwards. As we are
interested in identifying a critical wavenumber to associate with the number of
arms in the starburst fracture, we assume the contact-line radius has the
following representation $R(\theta)=R\left(1+\epsilon \cos n\theta\right)$, where
$n$ is the azimuthal wavenumber. 
We then apply this functional form for $R(\theta)$ to eqn.~(\ref{forceG}) and expand for small $\epsilon \ll 1$ to show the non-uniform $O(\epsilon)$ part of the wetting force is given by
\begin{equation}
F(r,\theta) = \left(-p H'(R-r)+ F_{CL} \delta'(r-R)\right)R\cos n\theta.
\label{appliedF}
\end{equation}
Equations~(\ref{navier}, \ref{dispBC}, \ref{forceBC} \&
\ref{appliedF}) define a boundary value problem governing the elastic
deformations of a substrate interacting with a liquid droplet with a corrugated
contact-line.

\subsection{Displacement potential}

We simplify the elastostatic governing equations (\ref{navier}) by introducing
the displacement potentials $\boldsymbol{G},\boldsymbol{A}$ defined such that
\begin{equation}
\boldsymbol{u} = \frac{1+\nu}{E} \left(2(1-\nu) \nabla^2
\boldsymbol{G} - \nabla\left(\nabla \cdot \boldsymbol{G}\right) + \nabla \times
\boldsymbol{A}\right),
\label{dispfunc}
\end{equation}
with
\begin{equation}
\boldsymbol{G} = \xi \left(r,\theta,z\right) \boldsymbol{\hat{e}}_z, \;
\boldsymbol{A} = 2 \psi(r,\theta,z) \boldsymbol{\hat{e}}_z.
\label{disppot}
\end{equation}
Substituting (\ref{dispfunc}) into the coupled system of differential equations
(\ref{navier}) delivers a set of uncoupled equations for $\xi,\psi$,
\begin{equation}
 \nabla^4 \xi=0, \; \nabla^2 \psi=0.
\label{fieldR}
\end{equation}
The displacement (\ref{dispBC}) and traction (\ref{forceBC})
boundary conditions can similarly be written in terms of the potential functions
$\xi,\psi$.

\subsection{Fourier-Hankel expansion}

We seek solutions to (\ref{fieldR}) for the potential functions using
Fourier-Hankel transforms. To begin, we assume $\xi,\psi$ has the following
Fourier decomposition,
\begin{equation}\begin{split}
\xi(r,\theta,z) = \sum_{n=0}^{\infty}{\xi_n(r,z) \cos
n\theta},\\ \psi(r,\theta,z) = \sum_{n=0}^{\infty}{\psi_n(r,z) \sin
n\theta}. \end{split}
\label{fourier}
\end{equation}
Next, the Hankel transform is applied to each Fourier component $\xi_n(r,z),
\psi_n(r,z)$ to yield
\begin{equation}\begin{split}
 \hat{\xi}_n (s,z) = \int_0^{\infty}{r \xi_n(r,z) J_n(sr) \mathrm{d}r},
\\ \hat{\psi}_n (s,z) = \int_0^{\infty}{r \psi_n(r,z) J_n(sr) \mathrm{d}r}, \end{split}
\label{hankel}
\end{equation}
where $J_n$ is the Bessel function of the first kind and $s$ is the radial
wavenumber. One applies the inverse Hankel transform
\begin{equation}\begin{split}
\xi_n (r,z) = \int_0^{\infty}{s \hat{\xi}_n(s,z) J_n(sr) \mathrm{d}s}, \\ \psi_n
(r,z) = \int_0^{\infty}{s \hat{\psi}_n(s,z) J_n(sr) \mathrm{d}s}, \end{split}
\label{invhankeldef}
\end{equation}
to recover $\xi_n, \psi_n$, and equivalently $\xi,\psi$.

\subsection{Reduced equations}
The following dimensionless variables are introduced:
\begin{equation}
u \equiv \hat{u} \frac{\sigma_{lg} \sin \alpha}{E}, ~~~
r \equiv \hat{r}h, ~~~
 z \equiv \hat{z}h, ~~~
 s \equiv \frac{\hat{s}}{h}, ~~~
 R \equiv \hat{R}h.
\label{scalings}
\end{equation}
with lengths  scaled by the thickness of the elastic substrate $h$ and
elastic deformations by the elastocapillary length $\ell \equiv
\frac{\sigma_{lg} \sin \alpha}{E}$. In the derivation that follows, we drop the
hats for notational simplicity. Substituting the expansions
(\ref{fourier}),(\ref{hankel}) into (\ref{fieldR}) gives reduced field equations
for $\hat{\xi}_n, \hat{\psi}_n$,
\begin{equation}\begin{split}
\nabla^4_n \hat{\xi}_n = \left( \frac{d^2}{dz^2}-s^2\right)^2
\hat{\xi}_n = 0, \\
\nabla^2_n \hat{\psi}_n = \left(\frac{d^2}{dz^2}-s^2\right)
\hat{\psi}_n = 0. \end{split}
\label{fieldRT}
\end{equation}

The following conditions are enforced on the rigid support $z=0$,
\begin{equation}
\frac{d\hat{\xi}_n}{dz}=0, \: \hat{\psi}_n =0, \: (1-2\nu)\frac{d^2 \hat{\xi}_n}{dz^2}-2(1-\nu)s^2\hat{\xi}_n = 0 ,\label{dispBCr2} \end{equation}
while on the free surface $z=1$ we require
\begin{widetext}
\begin{equation} \begin{split} & \nu \frac{d^3 \hat{\xi}_n}{dz^3}+(1-v) s^2 \frac{d \hat{\xi}_n}{dz} =0, \: \frac{d\hat{\psi}_n}{dz} = 0,\\ &
(1 - v)\frac{d^3 \hat{\xi}_n}{dz^3}- (2 - v)s^2 \frac{d\hat{\xi_n}}{dz} +
  \Upsilon (1 + v)s^2((1 - 2v)\frac{d^2 \hat{\xi}_n}{dz^2}- 2 (1 - v)s^2 \hat{\xi}_n) = \hat{F}_n, \end{split}\label{forceBCr2}\end{equation}
with
\begin{equation}
\hat{F}_n(s)=\Lambda \left((n-3)J_n(\Lambda s)-\Lambda s
J_{n-1}(\Lambda s)\right).
\end{equation}
The following dimensionless groups arise naturally from this choice of scaling,
\begin{equation}
\Upsilon\equiv \frac{\Sigma_{s}}{E h} ~~~ \mathrm{and} ~~~
\Lambda\equiv \frac{R}{h}.
\label{dimgroups}
\end{equation}
Here $\Upsilon$ is the solid elastocapillary number and $\Lambda$ is the aspect ratio
or dimensionless contact-line radius. The solution to
(\ref{fieldRT})--(\ref{forceBCr2}) is given by
\begin{equation}
\hat{\xi}_n = C_n \left(\cosh (sz) + \frac{s z \sinh (sz)}{2(1-2\nu)}\right)+
D_n \left(sz \cosh (sz) - \sinh (sz)  \right), \\ \hat{\psi}_n = 0,\;
\label{fieldsol}
\end{equation}
where
\begin{equation}\begin{split}
C_n=\frac{4
\hat{F}
_n(s)\left(2\nu-1\right)\left(s\cosh(s)+\left(2\nu-1\right)\sinh(s)\right)}{
s^3\left(5+2s^2-12\nu+8\nu^2+\left(3-4\nu\right)\cosh(2s)+2s\Upsilon\left(1-\nu^2
\right)\left(\left(3-4\nu\right)\sinh(2s)-2s\right)\right) }, \\
D_n=\frac{2
\hat{F}_n(s)\left(s\sinh(s)+2(1-\nu)\cosh(s)\right)}{
s^3\left(5+2s^2-12\nu+8\nu^2+\left(3-4\nu\right)\cosh(2s)+2s\Upsilon\left(1-\nu^2
\right)\left(\left(3-4\nu\right)\sinh(2s)-2s\right)\right) }.\end{split}
\label{CDdef}
\end{equation}

We compute the Fourier components $\xi_n,\psi_n$ in real space by evaluating the
inverse Hankel transform (\ref{invhankeldef}).
Once the potential functions $\xi,\psi$ are known, the displacement
$\boldsymbol{u}$, strain $\boldsymbol{\varepsilon}$ and stress
$\boldsymbol{\sigma}$ fields are obtained via substitution into
(\ref{dispfunc}), (\ref{kinematic}) and (\ref{constitutive}), respectively.

\end{widetext}

\subsection{Results}

We compute the elastic response in the substrate for a
fixed wavenumber $n$, consistent with the boundary conditions
(\ref{dispBC},\ref{forceBC}). That is, we assume that a corrugated
contact-line, with wavenumber $n$, gives rise to surface tractions
(\ref{forceBC}) that generate displacements $\boldsymbol{u}$, and equivalently
stresses $\boldsymbol{\sigma}$ and strains $\boldsymbol{\varepsilon}$, within
the gel substrate. We seek to develop a criteria based upon our solution for the elastic fields, that will yield a prediction for the critical wavenumber $n_c$ that is consistent with the number of fractures seen in experiments. This value can then be compared with the number of
fracture arms observed in experiments.

For the problem considered here, there are $3$ independent components of the displacement field, $6$ for the strain field and $6$ for the stress field, all of which vary in the two-dimensional $(r,z)$ space (domain) in cylindrical coordinates. We utilize several experimental observations to formulate our criteria. Firstly, we restrict our criteria to components of the displacement field, as failure in agarose gels typically correspond to disentanglement of the helices within the network of cross-linked, polymeric bonds. Secondly, experiments reveal that fractures are typically of the Mode $I$ (opening) variety \cite{daniels07}. These observations suggest our `failure criteria' should correlate with the maximum value of the $u_{\theta}$ component of the displacement field. Accordingly, we compute the displacement $u_\theta(r,z)$ for fixed values of the parameters $(\Lambda, \Upsilon, n, \nu)$. We simplify our problem further by setting the Poisson ratio $\nu=1/2$, which is a reasonable assumption for incompressible agarose gels.  In this case, the elastic modulus is then simply related to the shear modulus through the relationship $E=3G$.

\begin{figure}
\begin{center}
\includegraphics[width=\linewidth]{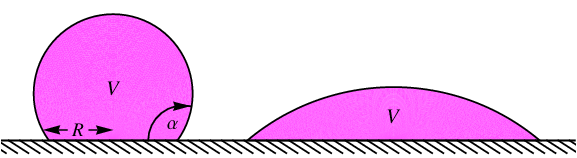}
\end{center}
\caption{\label{fig:SC} (Color online) Schematic of the relationship between the contact-line
radius $R$ and static contact-angle $\alpha$ for two spherical-cap droplets of
identical volume $V$. }
\end{figure}

The connection between the problem considered here and the experiment in
question is somewhat subtle. For a given set of experimental conditions, one
defines the liquid/solid (wetting) interactions by the choice of
fluid, which sets (equivalently) either the surface tension contrast $\Delta
\sigma$ or the static contact-angle through the relation
$\Delta\sigma/\sigma_{lg}= \cos\alpha$. Hence, the unbalanced contact-line force
$F_{CL}=\sigma_{lg} \sin\alpha$ could also be written in terms of the surface
tension contrast $\Delta\sigma$, if desired. Note that increasing the surface
tension contrast $\Delta\sigma$ is equivalent to decreasing the static
contact-angle $\alpha$ and consequently the magnitude of the contact-line force
$F_{CL}$. Naively, one might think that fixing the droplet volume $V$, as is
typically done in experiments, would set the contact-line radius $R$ and hence
the location of the contact-line force. However, $R$ is also a function of the
the wetting conditions. For a spherical-cap droplet, this relationship is given by
\begin{equation}
R^3=\frac{3V}{\pi}\left(\frac{\sin^3\alpha}{2-3\cos\alpha+\cos^3\alpha}\right),
\label{CLraddef}
 \end{equation}
as shown schematically in Figure~\ref{fig:SC}. The
difficulty is that adjusting $\Delta\sigma$ for fixed $V$, as is
typically done in experiments, changes both $R$ and $\alpha$. The benefit is that
it is possible to explore a wide range of contact-line radii $R$ by changing either $V$ or
$\Delta\sigma$.

\begin{figure*}
\begin{center}
\begin{tabular}{ccc}
($a$) & ($b$)  & ($c$) \\
\includegraphics[width=0.33\textwidth]{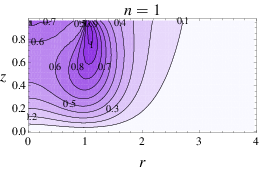} &
\includegraphics[width=0.33\textwidth]{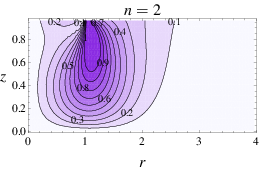} &
\includegraphics[width=0.33\textwidth]{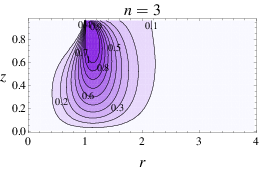}
\end{tabular}
\end{center}
\caption{\label{fig:hoopstrains0} (Color online) Tangential displacement $u_{\theta}$ in the
$(r,z)$ plane for wavenumber ($a$) $n=1$, ($b$) $n=2$ and ($c$) $n=3$ with surface tractions applied at contact-line radius $\Lambda=1$ with
$\Upsilon=0.01$. }
\end{figure*}

\begin{figure*}
\begin{center}
\begin{tabular}{ccc}
($a$) & ($b$)  & ($c$) \\
\includegraphics[width=0.33\textwidth]{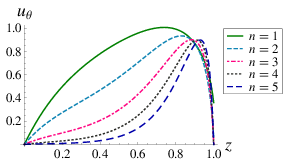} &
\includegraphics[width=0.33\textwidth]{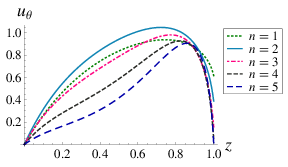} &
\includegraphics[width=0.33\textwidth]{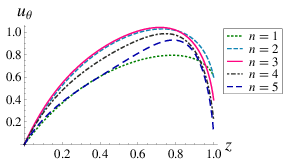}
\end{tabular}
\end{center}
\caption{\label{fig:hoopstrains} (Color online) Displacement $u_{\theta}$ against depth $z$
beneath the contact-line $r=R$ for ($a$) $\Lambda=1/2$, ($b$) $\Lambda=1$ and ($c$)
$\Lambda=3/2$, as it depends upon wavenumber $n$ for
$\Upsilon=0.01$. Solid line type denotes the critical disturbance $n_c$.}
\end{figure*}

To summarize, both the magnitude $F_{CL}$ and
location $R$ of the contact-line force implicitly depend upon the wetting
conditions through the static contact-angle $\alpha$. Henceforth, we restrict
ourselves to variations in $R$, and equivalently $\Lambda$. To isolate the
critical disturbance $n_c$, we fix the parameters $\Lambda, \Upsilon$ and
compute the displacement $u_{\theta}$ everywhere in the $(r,z)$ plane for each
wavenumber $n$ searching for its maximum value, our failure criteria.
As such, Figure~\ref{fig:hoopstrains0} shows that the location of maximum
displacement occurs directly beneath the contact-line $\Lambda=1$ at some finite
depth below the gel surface $z=1$.
For this set of parameters $\Lambda=1, \Upsilon=0.01$, the critical disturbance
is the $n=2$ mode, as it generates the greatest elastic response in the gel
substrate (see Figure~\ref{fig:hoopstrains}($a$)).

We systematically study the parametric dependence of the critical disturbance
upon the parameters $\Lambda, \Upsilon$ utilizing the observation that the
maximum displacement occurs directly beneath the contact-line.
Figure~\ref{fig:hoopstrains} plots the displacement $u_{\theta}$ as a function
of depth $z$ for various wavenumber $n$ disturbances to demonstrate that the
critical disturbance depends strongly upon the contact-line radius $\Lambda$.
For example, Figure~\ref{fig:hoopstrains}($a$) shows that the $n=1$ mode is the
critical disturbance for $\Lambda=1/2$, whereas the $n=3$ mode is the most
dangerous for $\Lambda=3/2$, as shown in Figure~\ref{fig:hoopstrains}($c$).
In general, one observes from Figure~\ref{fig:hoopstrains} that the critical
disturbance is the one that penetrates deepest into the elastic layer.

\begin{figure*}
\begin{center}
\begin{tabular}{cc}
($a$) & ($b$)  \\
\includegraphics[width=0.4\textwidth]{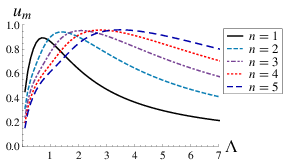} &
\includegraphics[width=0.4\textwidth]{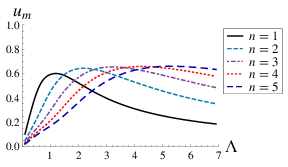}
\end{tabular}
\end{center}
\caption{\label{fig:max1} (Color online) Maximum displacement $u_m$ against aspect ratio
$\Lambda$ for ($a$) $\Upsilon=0.1$ and ($b$) $\Upsilon=1$.  }
\end{figure*}

\begin{figure}
\begin{center}
\includegraphics[width=0.4\textwidth]{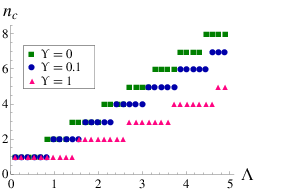}
\end{center}
\caption{\label{fig:dispersion1} (Color online) Critical wavenumber $n_c$ against aspect ratio
$\Lambda$, as it depends upon $\Upsilon$.  }
\end{figure}

\begin{figure}
\begin{center}
\includegraphics[width=0.4\textwidth]{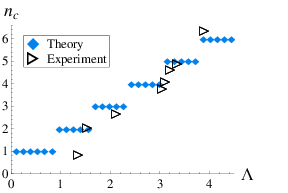}
\end{center}
\caption{\label{fig:dispersionEXPa} (Color online) Comparison to experiments from Daniels \emph{et al.} 2007 \cite{daniels07}: critical wavenumber $n_c$ against aspect ratio $\Lambda$ for $\Upsilon=0.1$. Each black triangle is the average of several experiments with $\Lambda$ (equivalently $R,\alpha$) estimated using reasonable values. }
\end{figure}

For fixed wavenumber $n$, we compute the maximum displacement $u_m$ through the
depth of the elastic layer as a function of contact-line radius $\Lambda$, as
shown in Figure~\ref{fig:max1} for two different solid elastocapillary numbers ($a$)
$\Upsilon=0.1$ and ($b$) $\Upsilon=1$.
For a given $\Lambda$, there is a wavenumber $n_c$ that generates the largest
elastic response, as measured by $u_\theta$. Here one should note the tight
spacing between curves in Figure~\ref{fig:max1}, which perhaps explains why a
range of wavenumbers are observed for a given experiment. More specifically,
slight variations in $\Lambda$ due to a number of experimental factors can alter
the expected wavenumber. Figure~\ref{fig:dispersion1} plots the critical
wavenumber $n_c$ against contact-line radius $\Lambda$. Here we note that our
theoretical prediction is consistent with experiments for fixed volume droplets
\cite{daniels07}, which indicate the number of arms in the starburst fracture
increases with increasing (decreasing) $\Delta \sigma$ ($\alpha$),
and equivalently increasing $\Lambda$. In fact, Figure~\ref{fig:dispersionEXPa} shows good quantitative agreement between theory and experiments for fixed volume droplets. For the comparison, we have estimated experimental $\Lambda$ values using reasonable values of the liquid/solid surface tension $\sigma_{ls}$ to compute the contact-angle $\alpha$ and equivalently $R$.

\begin{figure}
\begin{center}
\includegraphics[width=0.4\textwidth]{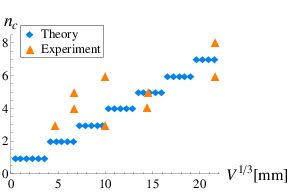}
\end{center}
\caption{\label{fig:dispersionEXP} (Color online) Critical wavenumber $n_c$ against $V^{1/3}$[mm]
for $\alpha=83^{\circ}$, $\Upsilon=0.1$ and $h=5$mm. Each orange triangle $\textcolor[rgb]{1.00,0.50,0.25}{\blacktriangle}$ is a single experiment.}
\end{figure}

Our model indicates that it is the \emph{location} $\Lambda$ of the contact-line force that selects the critical
wavenumber $n_c$, as opposed to the magnitude of that force. In fact, the
magnitude of the contact-line force, $F_{CL}/\sigma_{lg}= \sin\alpha$, decreases with increasing
(decreasing) $\Delta \sigma$ ($\alpha$), which is somewhat paradoxical
considering that naively one expects less fractures for smaller forces. Our
interpretation of the fracture mechanism is confirmed in
Figure~\ref{fig:dispersionEXP}, which plots the experimentally-observed
wavenumber against $V^{1/3}$ for fixed $\alpha$ compared to the model's
prediction. We observe quantitative agreement between the model and
experiments.

\begin{figure}
\begin{tabular}{c}
($a$)  \\
\includegraphics[width=0.4\textwidth]{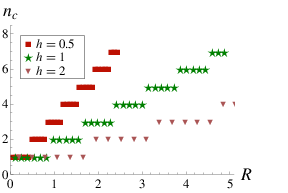}\\
($b$) \\
\includegraphics[width=0.5\textwidth]{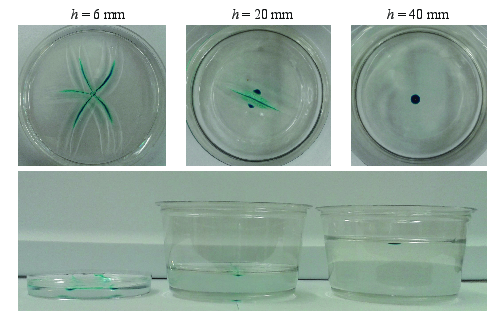} 
\end{tabular}
\caption{\label{fig:dispersionH} (Color online) Depth dependence: ($a$) critical wavenumber
$n_c$ against contact-line radius $R$ as a function of the height $h$ of the
elastic layer and ($b$) sample experiments showing typical results for droplets deposited on 0.08\%w agar substrate, with three different substrate thicknesses. }
\end{figure}

Our computations reveal that the thickness $h$ of the gel substrate is also a
critical parameter in determining the critical wavenumber, as $\Lambda\equiv
R/h$ couples two length scales, $R$ and $h$. Figure~\ref{fig:dispersionH} plots
the critical wavenumber $n_c$ against the unscaled contact-line radius $R$ for
various values of substrate height $h$. As shown, the critical wavenumber
increases (decreases) with decreasing (increasing) substrate height $h$. This
result, in effect, implies that large substrate heights are better able to
mitigate the imposed surface tractions. Experiments conducted on substrates of
two different thicknesses confirm this prediction. Figures~\ref{fig:dispersionH} ($b$) \& ($c$) are two experiments both conducted on substrates
with elastic modulus $E=16 $~Pa, but differing thickness ($b$)
$h=1.5$~cm, ($c$) $h=10$~cm and surface tension contrast ($b$) $\Delta \sigma=
17$~mN/m, ($c$) $\Delta \sigma=30$~mN/m, whose critical wavenumber is given by
($b$) $n=5$ and ($c$) $n=2$, respectively. If the
thickness of the elastic layer were unimportant, then one would expect
($c$) to produce more arms than ($b$), because of the greater surface tension
contrast. We observe the opposite, demonstrating that the thickness of the
elastic layer is more important in identifying the critical wavenumber.

\begin{figure}
\begin{center}
\begin{tabular}{c}
($a$) \\
\includegraphics[width=0.4\textwidth]{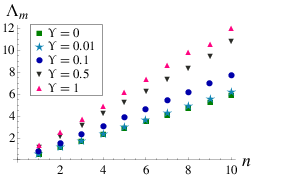} \\ ($b$) \\
\includegraphics[width=0.4\textwidth]{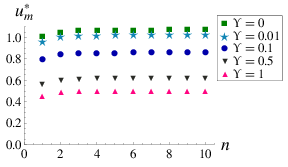}
\end{tabular}
\end{center}
\caption{\label{fig:max} (Color online) Maximal disturbance: ($a$) aspect ratio $\Lambda_m$ and
($b$) displacement $u_m$ against wavenumber $n$ as a function of elastocapillary
number $\Upsilon$.  }
\end{figure}

Lastly, we note that for a given wavenumber $n$ there is a preferred
contact-line radius $\Lambda_m$ that generates the largest elastic response
$u^*_m$, as shown in Figure~\ref{fig:max1}.  In Figure~\ref{fig:max}, we plot
the ($a$) location $\Lambda_m$ and ($b$) value $u^*_m$ of the maximal
disturbance as a function of wavenumber $n$, as it depends upon the
solid elastocapillary number $\Upsilon$. First, we note that the location $\Lambda_m$
is an increasing function of both wavenumber $n$ and solid elastocapillary number
$\Upsilon$. In contrast, Figure~\ref{fig:max} ($b$) shows that $u^*_m$ is a
slightly increasing function of $n$ that plateaus for large wavenumber. More
specifically, the increase in $u^*_m$ is particularly evident in
Figure~\ref{fig:max1} ($b$) when comparing the $n=1$ to the $n=2,3,\cdots$
disturbances. This implies that if, in experiment, one observes $n$ fractures,
then it is always possible to adjust the droplet volume $V$ to generate
$n-1,n-2,\cdots$ fractures. The converse is not necessarily true.
Finally, we show that $u^*_m$ decreases with $\Upsilon$ reflecting the energetic
penalty for flexure associated with the solid surface tension. In the
large-$\Upsilon$ regime, the gel substrate behaves more like a liquid than a
solid. We note that this result is not inconsistent with larger deformations for weaker substrates, as the real displacements are scaled with the elastocapillary length (\ref{scalings}). That is, decreasing the elastic modulus $E$ increases the solid elastocapillary number $\Upsilon$, but also the elastocapillary length $\ell$ and, thus, the real displacements.

We end this section with a comment on the failure criteria. The results presented in this section have assumed that the substrate is always weak enough to generate deformations that exceed the failure threshold for agar gel. As such, we have focused our study on identifying a critical wavenumber and not the failure threshold. Additional experiments would be needed to extract such information.


\section{Crack nucleation \label{sec:nucleation}}

The model developed above captures the central feature required to generate a starburst
with a particular number of arms: a critical wavenumber $n_c$ for which the tangential displacement is maximal. In experiments, we observe that for a given set of
experimental conditions, a narrow range of wavenumbers are selected; this is
not surprising given that values of $n \approx n_c$ can have a maximum
displacement similar to $n_c$. However, the failure criteria does not provide a failure mechanism. In physical
gels such as agar or gelatin, the stiffness of the material is provided by a
network of entangled polymers \citep{Goldbart1989a}. Under strain, these polymers
are stretched further away from equilibrium, although not necessarily to the
point of failure. Because polymers are subject to thermal fluctuations, even a
sub-critical strain can be triggered by a sufficiently large thermal
fluctuation. This is the mechanism behind delayed fracture, whereby there
is an exponential distribution of waiting times (indicating a Poisson process)
before which a strained material fractures. This has been directly
observed in both stiffer gelatin rods ($E=50$~kPa) \cite{bonn} and
in vibration-controlled experiments in this system. 


\section{Crack propagation \label{sec:propagation}}

\begin{figure}
\begin{center}
\includegraphics[width=0.37\textwidth]{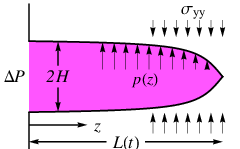}
\end{center}
\caption{\label{fig:fracture} (Color online) Definition sketch of a fluid-filled crack of
length $L$ driven by a pressure head $\Delta P$. }
\end{figure}

Once a fracture is initiated, fluid begins to flow into the crack tip and the crack propagates outward (away from the central droplet)
until the reservoir of fluid is exhausted. We develop a two-dimensional model for the asymptotic
propagation of a Mode I fracture driven by a pressure
head $\Delta P$. A schematic is shown in Figure~\ref{fig:fracture}. To begin, we
consider an incipient crack of length $L(t)$, filled with an inviscid fluid of
density $\rho$, propagating into an elastic solid.In this section, we model the solid as either i) a linear
elastic material or ii) an incompressible neo-Hookean material, as is typically
used for polymeric gels, and contrast the computed propagation rates.

We assume the capillary pressure head $\Delta P= 2 \sigma_{lg}/R$ at the fluid
reservoir $z=0$ is greater than the elastic stress in the solid in front of that
reservoir, so as to drive fluid into the crack tip. The flow $v$ of this
inviscid fluid is governed by Bernoulli's law,
\begin{equation} \frac{1}{2} \rho v^2 + p = C,
\label{bernoulli}
\end{equation}
and, hence, driven by the gradient in fluid pressure $p$, which is related to
the elastic stress induced by deformations of the crack surface. Here we assume
there is no ambient stress field in the elastic solid, or the substrate is not
pre-stressed. The localized stress (pressure) field near the crack tip
$z\rightarrow L(t)$ of a linear elastic solid has the well-known asymptotic form
\cite{pook},
\begin{equation}
\sigma_{yy}=-p=\frac{K_{I}}{\sqrt{2\pi}} \frac{1}{\sqrt{L(t)-z}}.
\label{Fstress}
\end{equation}

For an incompressible neo-Hookean solid, the asymptotic stress field is given by
\cite{stephenson82,krishnan09},
\begin{equation} \sigma_{yy}=-p =\frac{K^*_{I}}{\sqrt{2\pi}}
\frac{1}{\left(L(t)-x\right)^{2/3}}.
\label{NHstress}
\end{equation}
Here $K_I$ and $K^*_I$ are stress intensity factors for a linear elastic and
incompressible neo-Hookean solid, respectively. In general, $K_I$ is a function
of the crack geometry and far-field loading conditions, but we may simplify our
model by utilizing several experimental observations.

Recent experiments have demonstrated self-healing behavior of fluid-filled
cracks in soft substrates \cite{spandagos12a,spandagos12b}. This observation is
characteristic of stable, or transport-limited, crack growth. That is, the speed
of crack growth is limited by the extent to which the fluid can be supplied
(transported) to the crack tip. For reference, unstable crack growth would be
controlled by the speed of an elastic wave in the solid, a characteristic which
is not observed experimentally. Hereafter, we assume the crack propagates in the
marginal state $K_I=K_{Ic}$, where $K_{Ic}$ is the fracture toughness,
consistent with stable crack growth.

We apply Bernoulli's law (\ref{bernoulli}) using the crack tip (state 1)
and a point upstream (state 2) to derive an equation relating the crack-tip
velocity $v=\dot{L}$ to the pressure gradient in the linear elastic solid
(\ref{Fstress}):

\begin{equation} \dot{L} = \left(\frac{2}{\pi}\right)^{1/4}
\sqrt{\frac{K_{Ic}}{\rho}} \left(\frac{1}{L}\right)^{1/4},
\end{equation}
whose solution gives
\begin{equation}
L \propto \left(\frac{K_{Ic}}{\rho}\right)^{2/5} t^{4/5}.
\label{tipEQ}
\end{equation}
This predicts a characteristic exponent of $\beta=4/5$ for the growth of the arm.
A similar relationship is derived for an incompressible neo-Hookean solid by
applying the respective stress field (\ref{NHstress}) instead:
\begin{equation} \dot{L} = \left(\frac{2}{\pi}\right)^{1/4}
\sqrt{\frac{K^*_{Ic}}{\rho}} \left(\frac{1}{L}\right)^{1/3}.
\end{equation}
For this system, the characteristic exponent is instead $\beta=3/4$:
\begin{equation}
L \propto \left(\frac{K^*_{Ic}}{\rho}\right)^{3/8} t^{3/4},
\label{tipEQNH}
\end{equation}

Note that both constitutive models
predict power law behavior, but the neo-Hookean model is able to reproduce the
universal exponent $t^{3/4}$ observed experimentally \cite{daniels07}.
We attribute this result to the observation that the gel substrate is incompressible
and that the large scale deformation of  the soft solid
are more aptly described by the nonlinear neo-Hookean constitutive law.

\begin{figure}[!h]
\begin{center}
\includegraphics[width=0.4\textwidth]{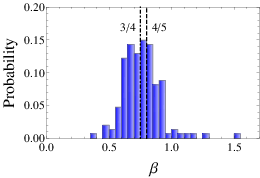}
\end{center}
\caption{\label{fig:alpha} (Color online) Histogram for the crack length exponent $\beta$ with
mean $\langle\beta\rangle=0.76$.  }
\end{figure}

Because the gels are spatially-heterogeneous, cracks nucleate with different widths and propagate at varying rates. We fit the length of each fracture arm to the form $L(t) = A t^\beta$ and examine the distribution of $\beta$ values observed. No systematic variation with material parameters was found for the values of $\beta$, but the width of the crack determines the pre-factor $A$ \citep{daniels07}. For this same data set of $147$ individual arms, we find that $\beta = 0.76 \pm 0.01$ (standard error). The full probability distribution is show in Figure~\ref{fig:alpha}, which allows for comparison to the $3/4$ (neo-Hookean) and $4/5$ (linear elastic) constitutive models.


\section{Discussion\label{sec:discussion}}
We have analyzed the deformations of a soft elastic substrate induced by the
liquid/solid interactions with a liquid droplet, as well as the
propagation of a fluid-filled crack in a soft elastic medium. The deformations associated with
a corrugated contact-line yield a critical
disturbance with wavenumber $n_c$ that generates the largest elastic response
within the substrate. We quantify the response with the tangential displacement, a failure
criteria which we correlate with the initiation of a starburst fracture (see Figure~\ref{fig:intro}).
Computations reveal that
the location of the unbalanced contact-line force is the most important
parameter in wavenumber selection. Our theoretical predictions compare favorably
to i) previously-reported experiments with fixed volume droplets \cite{daniels07} and ii) experiments where the contact-angle $\alpha$ is fixed and the droplet volume $V$ is varied,
thereby confirming our interpretation of the initiation process. Our model also predicts
that the substrate thickness $h$ is also an important parameter in wavenumber selection, an
observation which we see in experiments. For the crack propagation
problem, we develop a model by considering the transport of an inviscid
fluid into the crack tip of either a linear elastic or incompressible neo-Hookean solid. While
both elastic models yield power law growth, the neo-Hookean model predicts the crack
length grows with universal exponent $L\propto t^{3/4}$ and does not scale with any
material parameters, consistent
with experimental observations \cite{daniels07,spandagos12a,spandagos12b}.

Studies on fracture necessarily take place on materials which are strong enough to resist
fracture until the point when the measurement is made. However, many biological
materials are soft enough that conventional fracture measurements are not possible
because the material is too weak to support its own weight. The present study suggests
that by depositing a droplet of known size and wettability, it could be possible to use
starburst fractures to measure the fracture toughness of very soft materials so long as they
remain fully supported by the dish. Furthermore, because the growth rate of fractures is
sensitive to whether or not the material is linear-elastic, this could also provide a
new method for materials characterization.

\section*{Acknowledgements} The authors are grateful for support from the National Science Foundation under grant
number DMS-0968258, as well as NC State's Undergraduate Research Office, to Michael Shearer for valuable discussions, and to Mark Schillaci for preliminary experiments useful in formulating the model.




\begin{thebibliography}{10}

\bibitem{Roman2010}
B~Roman and J~Bico.
\newblock {Elasto-capillarity: deforming an elastic structure with a liquid
  droplet.}
\newblock {\em Journal of Physics: Condensed Matter}, 22(49):493101, November
  2010.

\bibitem{kajiya13}
Tadashi Kajiya, Adrian Daerr, Tetsuharu Narita, Laurent Royon, Francois
  Lequeux, and Laurent Limat.
\newblock Advancing liquid contact line on visco-elastic gel substrates:
  stick-slip vs. continuous motions.
\newblock {\em Soft Matter}, 9:454--461, 2013.

\bibitem{Snoeijer}
Antonin Marchand, Siddhartha Das, Jacco~H. Snoeijer, and Bruno Andreotti.
\newblock Contact angles on a soft solid: from {Young's Law to Neumann's Law}.
\newblock {\em Phys. Rev. Lett.}, 109:236101, Dec 2012.

\bibitem{Weijs2013}
Joost~H Weijs, Bruno Andreotti, and Jacco~H Snoeijer.
\newblock {Elasto-capillarity at the nanoscale : on the coupling between
  elasticity and surface energy in soft solids}.
\newblock {\em Soft Matter}, 9:8494--8503, 2013.

\bibitem{jerison11}
Elizabeth~R. Jerison, Ye~Xu, Larry~A. Wilen, and Eric~R. Dufresne.
\newblock Deformation of an elastic substrate by a three-phase contact line.
\newblock {\em Phys. Rev. Lett.}, 106:186103, May 2011.

\bibitem{style12}
Robert~W. Style and Eric~R. Dufresne.
\newblock Static wetting on deformable substrates{,} from liquids to soft
  solids.
\newblock {\em Soft Matter}, 8:7177--7184, 2012.

\bibitem{Das2011}
Siddhartha Das, Antonin Marchand, Bruno Andreotti, and Jacco~H. Snoeijer.
\newblock {Elastic deformation due to tangential capillary forces}.
\newblock {\em Phys. Fluids}, 23(7):072006, 2011.

\bibitem{huang07}
Jiangshui Huang, Megan Juszkiewicz, Wim~H. de~Jeu, Enrique Cerda, Todd Emrick,
  Narayanan Menon, and Thomas~P. Russell.
\newblock Capillary wrinkling of floating thin polymer films.
\newblock {\em Science}, 317(5838):650--653, 2007.

\bibitem{vella10}
D.~Vella, M.~Adda-Bedia, and E.~Cerda.
\newblock Capillary wrinkling of elastic membranes.
\newblock {\em Soft Matter}, 6:5778--5782, 2010.

\bibitem{Davidovitch2011}
Benny Davidovitch, Robert~D Schroll, Dominic Vella, Mokhtar Adda-Bedia, and
  Enrique Cerda.
\newblock {Prototypical model for tensional wrinkling in thin sheets.}
\newblock {\em Proceedings of the National Academy of Sciences}, 108(45),
  October 2011.

\bibitem{jung09}
Sunghwan Jung, Pedro~M. Reis, Jillian James, Christophe Clanet, and John W.~M.
  Bush.
\newblock Capillary origami in nature.
\newblock {\em Phys. Fluids}, 21(9):091110, 2009.

\bibitem{evans13}
Arthur~A. Evans, Saverio~E. Spagnolie, Denis Bartolo, and Eric Lauga.
\newblock Elastocapillary self-folding: buckling{,} wrinkling{,} and collapse
  of floating filaments.
\newblock {\em Soft Matter}, 9:1711--1720, 2013.

\bibitem{Mora2006}
T~Mora and a~Boudaoud.
\newblock {Buckling of swelling gels.}
\newblock {\em The European Physical Journal E: Soft matter}, 20(2):119--24,
  June 2006.

\bibitem{daniels07}
Karen~E. Daniels, Shomeek Mukhopadhyay, Paul~J. Houseworth, and Robert~P.
  Behringer.
\newblock Instabilities in droplets spreading on gels.
\newblock {\em Phys. Rev. Letters}, 99:124501, 2007.

\bibitem{spandagos12a}
Constantinos Spandagos, Thomas~B. Goudoulas, Paul~F. Luckham, and Omar~K.
  Matar.
\newblock Surface tension-induced gel fracture. part 1. fracture of agar gels.
\newblock {\em Langmuir}, 28(18):7197--7211, 2012.

\bibitem{spandagos12b}
Constantinos Spandagos, Thomas~B. Goudoulas, Paul~F. Luckham, and Omar~K.
  Matar.
\newblock Surface tension-induced gel fracture. part 2. fracture of gelatin
  gels.
\newblock {\em Langmuir}, 28(21):8017--8025, 2012.

\bibitem{bonn}
D~Bonn, H~Kellay, M~Prochnow, K~Ben-Djemiaa, and J~Meunier.
\newblock {Delayed Fracture Of An Inhomogeneous Soft Solid}.
\newblock {\em Science}, 280(5361):265--267, Apr 1998.

\bibitem{Wang2012a}
Xiao Wang and Wei Hong.
\newblock {Delayed fracture in gels}.
\newblock {\em Soft Matter}, 2012.

\bibitem{young}
T.~Young.
\newblock An essay on the cohesion of fluids.
\newblock {\em Phil. Trans. R. Soc. London}, 95:65--87, 1805.

\bibitem{dupre}
A.~Dupr\'{e}.
\newblock {\em Th\'{e}orie M\'{e}chanique de La Chaleur}.
\newblock Paris, Gauthier-Villars, Paris, France, 1869.

\bibitem{Goldbart1989a}
Paul Goldbart and Nigel Goldenfeld.
\newblock {Microscopic theory for cross-linked macromolecules. I. Broken
  symmetry, rigidity, and topology}.
\newblock {\em Physical Review A}, 39(3):1402--1411, feb 1989.

\bibitem{pook}
L.P. Pook.
\newblock {\em Linear Elastic Fracture Mechanics for Engineers: Theory and
  Applications}.
\newblock WIT Press, Boston, MA, 2000.

\bibitem{stephenson82}
R.A. Stephenson.
\newblock The equilibrium field near the tip of a crack for finite plane strain
  of incompressible elastic materials.
\newblock {\em J. Elasticity}, 2:65--99, 1982.

\bibitem{krishnan09}
V.R. Krishnan and C.-Y. Hui.
\newblock Finite strain stress fields near the tip of an interface crack
  between a soft incompressible elastic material and a rigid substrate.
\newblock {\em The European Physical Journal E}, 29:61--72, 2009.

\end{thebibliography}
\end{document}